\documentclass[twocolumn,showpacs,preprintnumbers]{revtex4}
\usepackage{dcolumn}
\usepackage{amssymb}
\usepackage{amsmath}
\usepackage{graphicx}
\usepackage{bm}

\begin{document}

\title{Dissipative relativistic fluid dynamics: a new way to derive the
equations of motion from kinetic theory}
\date{\today }
\author{G.S.\ Denicol${}^{a}$, T.\ Koide${}^{b}$, and D.H.\ Rischke${}^{a,b}$}
\affiliation{$^{a}$Institute f\"ur Theoretische Physik, Johann Wolfgang
Goethe-Universit\"at, Max-von-Laue Str.\ 1, D-60438 Frankfurt am Main,
Germany}
\affiliation{$^{b}$Frankfurt Institute for Advanced Studies, Johann Wolfgang
Goethe-Universit\"at, Ruth-Moufang Str.\ 1, D-60438 Frankfurt am Main,
Germany}

\begin{abstract}
We re-derive the equations of motion of dissipative relativistic fluid
dynamics from kinetic theory. In contrast to the derivation of Israel and
Stewart, which considered the second moment of the Boltzmann equation to
obtain equations of motion for the dissipative currents, we directly use the
latter's definition. Although the equations of motion obtained via the two
approaches are formally identical, the coefficients are different. We show
that, for the one-dimensional scaling expansion, our method is in better
agreement with the solution obtained from the Boltzmann equation.
\end{abstract}

\maketitle

Dissipative relativistic fluid dynamics is an effective theory to describe
the long-wavelength, low-frequency dynamics of various systems, with
important applications in relativistic heavy-ion collisions and astrophysics 
\cite{Pasi}. However, the derivation of dissipative relativistic fluid
dynamics from the underlying microscopic theory is not yet completely
established.

It can be rigorously shown that, in the \textit{non-relativistic, classical,
dilute-gas limit}, the Boltzmann equation becomes equivalent to the
microscopic BBGKY equations \cite{Liboff}. 
In the \textit{relativistic case\/} and/or for 
\textit{quantum fluids}, a rigorous proof does not exist, but it is commonly
assumed that the Boltzmann equation is a reliable approximation to the
underlying microscopic dynamics, in particular for dilute gases. Then, fluid
dynamics can be systematically derived by introducing an appropriate
coarse-graining scheme.

The Chapman-Enskog expansion \cite{chapman} is the most common method to
extract the fluid-dynamical equations of motion from the Boltzmann equation.
However, this method is not suitable for relativistic systems, since it will
inevitably lead to relativistic Navier-Stokes theory which displays
intrinsic problems such as acausality and instabilities \cite{his,dkkm4,pu}.

Israel and Stewart (IS) derived relativistic fluid-dynamical equations that
do not exhibit this problem, by extending the method proposed by Grad for
non-relativistic systems \cite{Grad}. In Grad's original work, the
single-particle distribution function is expanded around its local
equilibrium value in terms of a complete set of Hermite polynomials \cite%
{Grad_Hermite}. However, the generalization of Grad's approach to
relativistic systems is non-trivial, since it is not easy to find a suitable
set of orthogonal polynomials which could replace the Hermite polynomials 
\cite{DeGroot,Stewart_Review}. Thus Israel and Stewart introduced another
approximation, the so-called 14-moment approximation \cite{IS}, where the
distribution function is expanded as a Taylor series in momentum around its
local equilibrium value. The expansion is truncated at second order in
momentum and only 14 coefficients remain to describe the distribution
function.

It is important to note that the derivation of Israel and Stewart contains
one additional approximation besides the 14 moments $\mathit{ansatz}$: they
used the second moment of the Boltzmann equation to extract the equations of
motion for the dissipative currents and, hence, to determine the transport
coefficients \cite{IS,Stewart_Review,DeGroot}. However, this choice to
extract the equations of motion is ambiguous, because \textit{any\/} moment
of the Boltzmann equation will lead to a closed set of equations, once the
14-moment approximation is applied. The transport coefficients appearing in
the final equations depend on the choice of the moment.

Thus, the choice of the moment is quite an important issue. In fact, it was
confirmed that, at least for some cases, the IS equations are not in good
agreement with the numerical solution of the Boltzmann equation \cite%
{denes,el}. Also, the transport coefficients obtained by Israel and Stewart
do not coincide with quantum-field theoretical calculations \cite{dhkr}.
These inconsistencies may arise because of an inappropriate choice of the
moment equation.

Then, which moment should be used to derive fluid dynamics? Remember that we
are interested in the equations of motion for the dissipative currents.
Furthermore, the dissipative currents are well-defined in terms of the
single-particle distribution function from the kinetic point of view.
Therefore, we can calculate the equations of motion for all the dissipative
currents directly from their definitions without referring to an arbitrary
moment of the Boltzmann equation. The purpose of this letter is to derive
new fluid-dynamical equations following this idea. We shall show that the
form of these equations is the same as in IS theory, but the values of the
coefficients are different. For the one-dimensional scaling expansion, we
demonstrate that the new equations agree better with a numerical solution of
the Boltzmann equation than the IS equations.

We start from the relativistic Boltzmann equation 
\begin{equation} \label{BE}
K^{\mu }\partial _{\mu }f_{K}=C\left[ f\right] ,
\end{equation}%
where $K^{\mu }=(E_{\mathbf{k}},\mathbf{k})$ with $E_{\mathbf{k}}=\sqrt{%
\mathbf{k}^{2}+m^{2}}$ with $m$ being the particle mass. In the collision
term we consider only elastic two-to-two collisions, 
\begin{eqnarray}
C\left[ f\right] &=&\frac{1}{2}\int dK^{\prime }dPdP^{\prime }\,W_{KK\prime
\rightarrow PP\prime }  \notag \\
&&\times \left( f_{P}f_{P^{\prime }}\tilde{f}_{K}\tilde{f}_{K^{\prime
}}-f_{K}f_{K^{\prime }}\tilde{f}_{P}\tilde{f}_{P^{\prime }}\right) \;.
\end{eqnarray}%
Here, $dK \equiv g\,d^{3}\vec{K}/\left[ (2\pi)^{3}E_{\mathbf{k}}\right] $ is
the Lorentz-invariant measure, with $g$ being the degeneracy factor, and $%
W_{KK\prime \rightarrow PP\prime }$ is the transition rate of the collision.
We used the notation $f_{K}\equiv f(x^{\mu},K^{\mu })$ and $\tilde{f}%
_{K}\equiv 1-af(x^{\mu },K^{\mu })$, where $a=1$ ($a=-1$) for fermions
(bosons) and $a=0$ for a Boltzmann gas.

The conserved particle current $N^{\mu }$ and the energy-momentum tensor $%
T^{\mu \nu }$ are expressed in terms of the single-particle distribution
function as 
\begin{align}
N^{\mu }& =\left\langle K^{\mu }\right\rangle \;, \\
T^{\mu \nu }& =\left\langle K^{\mu }K^{\nu }\right\rangle \;,
\end{align}%
where $\left\langle \ldots \right\rangle \equiv \int dK\left( \ldots \right)
f_{K}$.

We introduce the fluid four-velocity $u^{\mu }$ as an eigenvector of the
energy-momentum tensor, $T^{\mu \nu }u_{\mu }=\varepsilon u^{\nu }$, where
the eigenvalue $\varepsilon $ is the energy density \cite{Landau}. Then, we
can decompose the four-momentum as 
\begin{equation}
K^{\mu }=\,\left( u\cdot K\right) u^{\mu }+K^{\left\langle \mu \right\rangle
}\,.
\end{equation}%
Here, we defined the scalar product of two four-vectors $A^{\mu },\,B^{\mu }$
as $A_{\mu }B^{\mu }\equiv A\cdot B$ and we introduced the projection
operator $\Delta ^{\mu \nu }=g^{\mu \nu }-u^{\mu }u^{\nu }$ and $%
A^{\left\langle \mu \right\rangle }=\Delta ^{\mu \nu }A_{\nu }$ for an
arbitrary four-vector $A^{\mu }$. The metric tensor is $g^{\mu \nu} \equiv 
\mathrm{diag}(+,-,-,-)$.

Using this decomposition, $N^{\mu }$ and $T^{\mu \nu }$ can be written in
the form, 
\begin{eqnarray}
N^{\mu } &=&nu^{\mu }+n^{\mu },  \notag \\
T^{\mu \nu } &=&\varepsilon \,u^{\mu }u^{\nu }-\Delta ^{\mu \nu }\left(
P+\Pi \right) +\pi ^{\mu \nu }\;,
\end{eqnarray}%
where the particle density $n$, the particle diffusion current $n^{\mu }$,
the energy density $\varepsilon $, the shear stress tensor $\pi ^{\mu \nu }$%
, and the sum of thermodynamic pressure, $P_{0}$, and bulk viscous pressure, 
$\Pi $, are defined by 
\begin{eqnarray}
n &\equiv &\left\langle u\cdot K\right\rangle \,,\;n^{\mu }\equiv
\left\langle K^{\left\langle \mu \right\rangle }\right\rangle
\,,\;\varepsilon \equiv \left\langle (u\cdot K)^{2}\right\rangle \,,  \notag
\\
\pi ^{\mu \nu } &\equiv &\left\langle K^{\left\langle \mu \right. }K^{\left.
\nu \right\rangle }\right\rangle \,,\;P_{0}+\Pi \equiv -\frac{1}{3}%
\left\langle \Delta ^{\mu \nu }K_{\mu }K_{\nu }\right\rangle \;,
\label{def_hy_qua}
\end{eqnarray}
where $A^{\langle \mu \nu \rangle }\equiv \Delta ^{\mu \nu \alpha \beta
}A_{\alpha \beta }$, with $\Delta ^{\mu \nu \alpha \beta }\equiv
(\Delta^{\mu \alpha }\Delta ^{\beta \nu }+\Delta ^{\nu \alpha }
\Delta^{\beta \mu }-\frac{2}{3}\Delta ^{\mu \nu }\Delta ^{\alpha \beta })/2$%
. We define the local equilibrium distribution function as $f_{0K}=\left(
\exp \left( \beta _{0}\,u\cdot K-\alpha _{0}\right) +a\right) ^{-1}$, where $%
\beta _{0}$ and $\alpha _{0}$ are the inverse temperature and the ratio of
the chemical potential to temperature, respectively. These are defined by
the matching conditions 
\begin{equation}
n\equiv n_{0}=\langle u\cdot K\rangle _{0}\,,\varepsilon \equiv \varepsilon
_{0}=\left\langle (u\cdot K)^{2}\right\rangle _{0},
\end{equation}
where $\langle \ldots \rangle _{0}\equiv \int dK\left( \ldots \right) f_{0K}$%
.

The separation between thermodynamic pressure and bulk viscous pressure is
then achieved by 
\begin{equation}
P_{0}=-\frac{1}{3}\,\langle \Delta ^{\mu \nu }K_{\mu }K_{\nu }\rangle
_{0}\,,\Pi =-\frac{1}{3}\,\left\langle \Delta ^{\mu \nu }K_{\mu }K_{\nu
}\right\rangle _{\delta }\;,
\end{equation}
with $\left\langle \ldots \right\rangle _{\delta }\equiv \left\langle \ldots
\right\rangle -\left\langle \ldots \right\rangle _{0}$.

So far, there is no difference between the calculation of Israel and Stewart
and ours. The difference emerges in the derivation of the equations of
motion for the dissipative currents. Israel and Stewart obtained these
equations from the second moment of the Boltzmann equation \cite%
{IS,Stewart_Review} 
\begin{equation}
\partial _{\mu }\left\langle K^{\mu }K^{\nu }K^{\lambda }\right\rangle =\int
dK\,K^{\nu }K^{\lambda }\,C\left[ f\right]\, .
\end{equation}
Then, the equations of motion for $\Pi $, $n^{\mu }$, and $\pi^{\mu \nu }$
are obtained by the projections $u_{\nu}u_{\lambda }\partial _{\mu }
\left\langle K^{\mu }K^{\nu }K^{\lambda}\right\rangle $, $\Delta _{\lambda
}^{\alpha }u_{\nu }\partial _{\mu }\left\langle K^{\mu }K^{\nu }K^{\lambda
}\right\rangle $, and $\Delta _{\nu \lambda }^{\alpha \beta }\partial_{\mu }
\left\langle K^{\mu }K^{\nu }K^{\lambda }\right\rangle $, respectively,
together with the 14-moment approximation for the single-particle
distribution function (see below). These equations determine the time
evolution of $\Pi $, $n^{\mu }$, and $\pi^{\mu \nu }$ through their comoving
derivatives, $\dot{\Pi}$, $\dot{q}^{\langle \mu \rangle} \equiv \Delta^{\mu
\nu} \dot{q}_\nu$, and $\dot{\pi}^{\langle \mu \nu \rangle} \equiv
\Delta^{\mu \nu \alpha \beta} \dot{\pi}_{\alpha \beta} $, respectively,
where $\dot{A} \equiv u \cdot \partial A$ is the comoving derivative.

However, we can calculate these comoving derivatives also directly from Eq.\
(\ref{def_hy_qua}): 
\begin{eqnarray}
\dot{\Pi} &=&-\frac{1}{3}m^{2}\int dK \, \delta \dot{f}\;,  \label{Exact1} \\
\dot{n}^{\left\langle \mu \right\rangle } &=&\int dK\, K^{\left\langle \mu
\right\rangle } \, \delta \dot{f}\;,  \label{Exact2} \\
\dot{\pi}^{\left\langle \mu \nu \right\rangle } &=&\int dK\, K^{\left\langle
\mu \right. }K^{\left. \nu \right\rangle }\, \delta \dot{f}\;.
\label{Exact3}
\end{eqnarray}
Then, using the Boltzmann equation (\ref{BE}) in the form
\begin{equation}
\delta \dot{f} = - \dot{f}_0 - \frac{1}{u \cdot K}\, K \cdot
\nabla f + \frac{1}{u \cdot K} C[f]\;,
\end{equation}
where $\nabla ^{\mu }\equiv \Delta^{\mu \nu }\partial_{\nu }$, 
we obtain the \textit{exact} equations 
\begin{eqnarray}
\dot{\Pi} &=&-C-\beta _{\Pi }\theta -\zeta _{\Pi \Pi }\Pi \theta +\zeta
_{\Pi \pi }\pi ^{\mu \nu }\sigma _{\mu \nu }  \notag \\
&&-\zeta _{\Pi n}\partial \cdot n+\frac{m^{2}}{3}\nabla _{\nu }\left\langle
(u\cdot K)^{-1}K^{\left\langle \nu \right\rangle }\right\rangle _{\delta } 
\notag \\
&&+\frac{m^{2}}{3}\left\langle (u\cdot K)^{-2}K^{\mu }K^{\nu }\right\rangle
_{\delta }\nabla _{\mu }u_{\nu },  \label{Exact11} \\
\dot{n}^{\left\langle \mu \right\rangle } &=&C^{\mu }+\beta _{n}\nabla ^{\mu
}\alpha _{0}-n^{\mu }\theta -n\cdot \nabla u^{\mu }  \notag \\
&&+h_{0}\left( \Pi \dot{u}^{\mu }-\nabla ^{\mu }\Pi \right) +h_{0}\Delta
^{\mu \nu }\partial _{\lambda }\pi _{\nu }^{\lambda }  \notag \\
&&-\Delta _{\nu }^{\mu }\nabla _{\alpha }\left\langle (u\cdot
K)^{-1}K^{\left\langle \nu \right\rangle }K^{\left\langle \alpha
\right\rangle }\right\rangle _{\delta }  \notag \\
&&-\left\langle (u\cdot K)^{-2}K^{\left\langle \mu \right\rangle }K^{\alpha
}K^{\beta }\right\rangle _{\delta }\nabla _{\alpha }u_{\beta },
\label{Exact22} \\
\dot{\pi}^{\left\langle \mu \nu \right\rangle } &=&C^{\mu \nu }+2\beta _{\pi
}\sigma ^{\mu \nu }-\frac{5}{3}\pi ^{\mu \nu }\theta  \notag \\
&&-2\pi _{\rho }^{\left\langle \mu \right. }\sigma ^{\left. \nu
\right\rangle \rho }+2\pi _{\rho }^{\left\langle \mu \right. }\omega
^{\left. \nu \right\rangle \rho }+2\Pi \sigma ^{\mu \nu }  \notag \\
&&-\Delta _{\lambda \sigma }^{\mu \nu }\nabla _{\rho }\left\langle (u\cdot
K)^{-1}K^{\left\langle \lambda \right\rangle }K^{\left\langle \sigma
\right\rangle }K^{\left\langle \rho \right\rangle }\right\rangle _{\delta } 
\notag \\
&&-\left\langle (u\cdot K)^{-2}K^{\left\langle \mu \right. }K^{\left. \nu
\right\rangle }K^{\alpha }K^{\beta }\right\rangle _{\delta }\nabla _{\alpha
}u_{\beta },  \label{Exact333}
\end{eqnarray}
where $h_{0}=n_{0}/(\varepsilon _{0}+P_{0})$, 
and we introduced the vorticity $\omega
_{\lambda \rho }\equiv \frac{1}{2}\left( \nabla _{\lambda }u_{\rho }-\nabla
_{\rho }u_{\lambda }\right) $, the shear tensor $\sigma _{\lambda \rho
}\equiv \nabla _{\left\langle \lambda \right. }u_{\left. \rho \right\rangle }
$ and the expansion scalar $\theta \equiv \nabla _{\mu }u^{\mu }$. Above, we
used the following notation for the collision terms, 
\begin{eqnarray}
C &=&\frac{m^{2}}{3}\int dK\left( u\cdot K\right) ^{-1}C\left[ f\right]\; , 
\notag \\
C^{\mu } &=&\int dK\left( u\cdot K\right) ^{-1}K^{\langle\mu \rangle}
C\left[ f\right]\; ,
\notag \\
C^{\mu \nu } &=&\int dK\left( u\cdot K\right) ^{-1}K^{\langle\mu }
K^{\nu \rangle}C\left[f\right]\; .
\end{eqnarray}

However, because the remaining terms in angular brackets cannot be entirely
expressed in terms of the macroscopic variables (\ref{def_hy_qua}), Eqs.\ (%
\ref{Exact11}), (\ref{Exact22}), and (\ref{Exact333}) are not closed. In
order to obtain a closed set of equations, we use the 14-moment
approximation for the single-particle distribution function introduced by
Israel and Stewart 
\begin{equation}
f_{K}=f_{0K}+f_{0K}\tilde{f}_{0K}\left( \lambda _{\Pi }\Pi +\lambda
_{n}n_{\alpha }K^{\alpha }+\lambda _{\pi }\pi _{\alpha \beta }K^{\alpha
}K^{\beta }\right)\; ,  \label{14m}
\end{equation}
and insert this into Eqs.\ (\ref{Exact11}), (\ref{Exact22}), and (\ref%
{Exact333}) to compute the terms in angular brackets. This system of
equations is now closed, since the approximation (\ref{14m}) solely involves
the quantities of Eq.\ (\ref{def_hy_qua}). The coefficients $\lambda _{\Pi }$%
, $\lambda _{n}$ and $\lambda _{\pi }$ are well-known
functions of $u \cdot K,\, \alpha_0,$ and $\beta_0$, see e.g.\ Refs.\ \cite%
{IS,Stewart_Review,DeGroot} for details.

We finally obtain the equations of dissipative relativistic fluid dynamics, 
\begin{eqnarray}
\dot{\Pi} &=&-\frac{\Pi }{\tau _{\Pi }}-\beta _{\Pi }\theta -\ell _{\Pi
n}\partial \cdot n-\tau _{\Pi n}n\cdot \dot{u}-\delta _{\Pi \Pi }\Pi \theta 
\notag \\
&&-\lambda _{\Pi n}n\cdot \nabla \alpha _{0}+\lambda _{\Pi \pi }\pi ^{\mu
\nu }\sigma _{\mu \nu }\;,  \label{eq_bulk} \\
\dot{n}^{\left\langle \mu \right\rangle } &=&-\frac{n^{\mu }}{\tau _{n}}%
+\beta _{n}\nabla ^{\mu }\alpha _{0}-n_{\nu }\omega ^{\nu \mu }-\delta
_{nn}n^{\mu }\theta -\ell _{n\Pi }\nabla ^{\mu }\Pi   \notag \\
&&+\ell _{n\pi }\Delta ^{\mu \nu }\partial _{\lambda }\pi _{\nu }^{\lambda
}+\tau _{n\Pi }\Pi \dot{u}^{\mu }-\tau _{n\pi }\pi _{\nu }^{\mu }\dot{u}%
^{\nu }  \notag \\
&&-\lambda _{nn}n^{\nu }\sigma _{\nu }^{\mu }+\lambda _{n\Pi }\Pi \nabla
^{\mu }\alpha _{0}-\lambda _{n\pi }\pi ^{\mu \nu }\nabla _{\nu }\alpha
_{0}\;,  \label{eq_heat} \\
\dot{\pi}^{\left\langle \mu \nu \right\rangle } &=&-\frac{\pi ^{\mu \nu }}{%
\tau _{\pi }}+2\beta _{\pi }\sigma ^{\mu \nu }+2\pi _{\alpha }^{\left\langle
\mu \right. }\omega ^{\left. \nu \right\rangle \alpha }-\tau _{\pi
n}n^{\left\langle \mu \right. }\dot{u}^{\left. \nu \right\rangle }  \notag \\
&&+\ell _{\pi n}\nabla ^{\left\langle \mu \right. }n^{\left. \nu
\right\rangle }-\delta _{\pi \pi }\pi ^{\mu \nu }\theta -\tau _{\pi \pi }\pi
_{\alpha }^{\left\langle \mu \right. }\sigma ^{\left. \nu \right\rangle
\alpha }  \notag \\
&&+\lambda _{\pi n}n^{\left\langle \mu \right. }\nabla ^{\left. \nu
\right\rangle }\alpha _{0}+\lambda _{\pi \Pi }\Pi \sigma ^{\mu \nu }\;.
\label{eq_shear}
\end{eqnarray}%
The derived equations contain 25 transport coefficients, of which we only
show the following three coefficients explicitly, 
\begin{eqnarray}
\beta _{\Pi } &=&\left( \frac{1}{3}-c_{s}^{2}\right) \left( \varepsilon
_{0}+P_{0}\right) -\frac{2}{9}\left( \varepsilon _{0}-3P_{0}\right)   \notag
\\
&&-\frac{m^{4}}{9}\langle \left( u\cdot K\right) ^{-2}\rangle _{0}, \\
\beta _{n} &=&\frac{2}{3\beta _{0}}\langle 1\rangle _{0}+\frac{m^{2}}{3\beta
_{0}}\langle \left( u\cdot K\right) ^{-2}\rangle _{0}-\frac{n_{0}}{\beta _{0}%
}h_{0}, \\
\beta _{\pi } &=&\frac{4}{5}P_{0}+\frac{1}{15}\left( \varepsilon
_{0}-3P_{0}\right) -\frac{m^{4}}{15}\langle \left( u\cdot K\right)
^{-2}\rangle _{0},
\end{eqnarray}%
where the velocity of sound (squared) is
$c_{s}^{2}=(dP_{0}/d\varepsilon _{0})_{s_{0}/n_0}$ where $s_0$ is
the entropy density. The other coefficients will be
reported in Ref.\ \cite{dkr}. While the form of the derived equations (\ref%
{eq_bulk}), (\ref{eq_heat}), and (\ref{eq_shear}) are the same as those
obtained in previous calculations \cite{BHR,Harri}, the transport
coefficients are different. That is, the derivation of the equations of
dissipative relativistic fluid dynamics from the Boltzmann equation is
ambiguous and depends on the method applied. We remark that, in the
non-relativistic (low-temperature) limit, the set of transport coefficients
as computed with the method of Israel and Stewart and ours converge to the
same values.

Now we would like to quantify the difference between the IS equations and
ours at hand of a simple example. We consider a massless Boltzmann gas
equation of state and the one-dimensional Bjorken scaling expansion, where
the velocity is given by $u^{\mu }=\frac{1}{\tau }\left( t,0,0,z\right) $,
and the fluid-dynamical variables are only a function of the proper time $\tau 
$. Then, $\Pi $ and $n^{\mu }$ vanish. The shear stress tensor $\pi ^{\mu
\nu }$ has only diagonal components and can be characterized by a function $%
\pi $ as $\pi ^{\mu \nu }= \mathrm{diag} \left( 0,\pi /2,\pi \,/2,-\pi
\right) $.

The equation for $\pi $ is given by 
\begin{equation}
\frac{d\pi }{d\tau }+\frac{\pi }{\tau _{\pi }}=\beta _{\pi }\frac{4}{3\tau }%
-\lambda \frac{\pi }{\tau }.  \label{shear_Bjorken}
\end{equation}%
In the massless limit, our transport coefficients simplify, 
\begin{equation}
\beta _{\pi }=\frac{4P_{0}}{5},~~\tau _{\pi }^{-1}=\frac{3}{5}%
\,\sigma P_{0}\beta_0 ,\text{ }\lambda \equiv \frac{4}{9}\tau _{\pi \pi
}+\delta _{\pi \pi }= \frac{124}{63}\;,  \label{tra_our}
\end{equation}
where $\sigma$ is the total cross section \cite{Carsten}. Here we
assumed that $\sigma$ is independent of energy and momentum as is done
in Refs.\ \cite{denes,el}. As mentioned above, the form of Eq.\ (\ref%
{tra_our}) is identical to that of IS theory, but the transport coefficients
assume different values. In IS theory, these coefficients are given by 
\begin{equation}
\beta _{\pi }=\frac{2P_{0}}{3},~~\tau _{\pi }^{-1}=\frac{5}{9}\,
\sigma P_{0}\beta_0 ,\text{ }\lambda =2\;.
\end{equation}
Equation (\ref{shear_Bjorken}) couples to the equation of the pressure which
is given by 
\begin{equation}
\frac{dP_{0}}{d\tau }+\frac{4P_{0}}{3\tau }-\frac{\pi }{3\tau }=0\;.
\label{energy}
\end{equation}

\begin{figure}[tbp]
\includegraphics[scale=0.4]{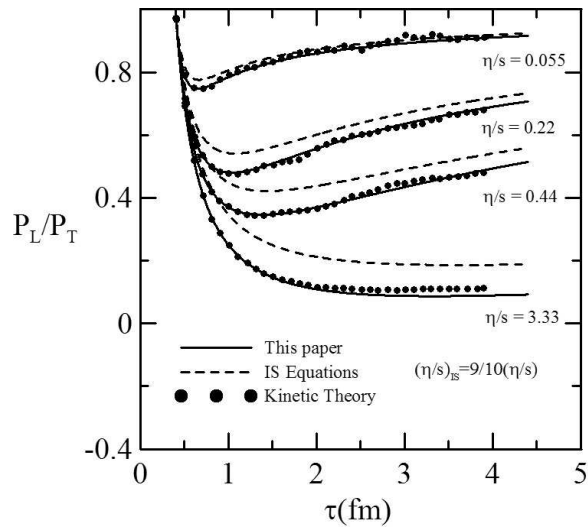}
\caption{ Time evolution of the ratio $P_{L}/P_T$ for our equations (solid
line), the IS equations (dotted line), and the numerical solution of the
Boltzmann equation (circles). }
\label{Bjorken_Comp}
\end{figure}

In Fig.\ \ref{Bjorken_Comp}, we show the time dependence of the anisotropy
of the effective pressure, which is defined by 
\begin{equation}
\frac{P_{L}(\tau )}{P_{T}(\tau )}=\frac{P_{0}(\tau )-\pi(\tau)}{ P_{0}(\tau)
+\pi (\tau )/2}.
\end{equation}
We used $T=500$ MeV and $\pi =0$ as initial condition. The solid and dashed
lines represent our result and the result of IS theory, respectively. The
circles correspond to the numerical solution of the Boltzmann equation \cite%
{el}. This calculation is performed with values for the cross section such
that the shear viscosity $\eta \equiv \beta_\pi \tau_\pi$ to entropy density 
$s$ ratio is constant. Since all the results must be compared by fixing a
common cross section, $\eta = 4/(3\sigma \beta_0)$ and $\eta_{IS} =
6/(5\sigma \beta_0)$ (the shear viscosity of the IS theory) have different
values and are related by $\eta =\frac{10}{9}\eta _{IS}$.

One can see that IS theory (the complete IS equations \cite{Harri}) always
overestimates the anisotropy obtained by the numerical solution of the
Boltzmann equation, even for very low viscosities ($\eta _{IS}/s=0.05$). On
the other hand, our equations clearly show a better agreement. Visible
deviations are only observed for the case of $\eta /s=3.33$ at late times.
This result indicates that our fluid-dynamical approach is better adapted
than IS theory to capture the microphysics contained in the Boltzmann
equation.

In summary, we have proposed a new method for deriving the fluid-dynamical
equations from kinetic theory. In our approach, the equations for the
dissipative currents are obtained directly from the definitions of these
currents. This method is different from the traditional IS approach \cite{IS}%
, where the equations are extracted from the second moment of the Boltzmann
equation. Our method can successfully reproduce the numerical solution of
the Boltzmann equation for the simple one-dimensional scaling expansion. It
is also important to mention that the transport coefficients of our kinetic
calculation are consistent with those calculated from quantum field theory
with the method proposed in Ref.\ \cite{dhkr}.

The authors thank A.\ El, C.\ Greiner, H.\ Niemi, P.\ Huovinen, and T.\
Kodama for fruitful discussions and their interest in this work. T.K.\
acknowledges inspiring discussions with T.\ Hatsuda, T.\ Hirano, and A.\ Monnai.
This work was supported by the Helmholtz International Center for FAIR
within the framework of the LOEWE program launched by the State of Hesse.

\end{document}